\documentclass[aps,prd,twocolumn,superscriptaddress,showpacs,nofootinbib,altaffilletter]{revtex4}
\usepackage{amsmath,amssymb}\begin{document}
\title{Singular fate of the universe in modified theories of gravity}
\author{L. Fern\'andez-Jambrina}
\email[]{leonardo.fernandez@upm.es}
\homepage[]{http://debin.etsin.upm.es/ilfj.htm}
\affiliation{Matem\'atica Aplicada, E.T.S.I. Navales, Universidad
Polit\'ecnica de Madrid,\\
Arco de la Victoria s/n, \\ E-28040 Madrid, Spain}%
\author{Ruth Lazkoz}\email[]{ruth.lazkoz@ehu.es}
\homepage[]{http://tp.lc.ehu.es/rls.html}\affiliation{F\'\i
sica Te\'orica, Facultad de Ciencia y Tecnolog\'\i a, Universidad del
Pa\'\i s Vasco,\\ Apdo.  644, E-48080 Bilbao, Spain}
\date{\today}
\begin{abstract}In this paper we study the final fate of the universe
in modified theories of gravity. As compared with general
relativistic formulations, in these scenarios the Friedmann equation
has  additional terms
which are relevant for low density epochs. We analyze the sort of
future singularities to be found  under the usual assumption the
expanding Universe is solely filled with a pressureless component.
We report our results using two schemes: one concerned with the
behavior of curvature scalars, and
a more refined one linked to observers. Some examples with a very
solid theoretical motivation
and some others with a more phenomenological nature are used for
illustration.
\end{abstract}

\maketitle

\section{Introduction}
Refined astronomical observations of luminosity distances derived
from Type Ia supernovae  provide reliable evidence of the current
cosmic speed up of the Universe
(see \cite{snpion} for the pioneering results and
\cite{Davis:2007na,WoodVasey:2007jb} for the latest).
In fact, such measurements are the only direct indication of that
phenomenon (see for e.g. \cite{Leibundgut:2004}), but at the same
time they are complementary with other key observations such as those
of the CMB spectrum
and the global matter distribution.
Explaining this surprising behavior in the large-scale
evolution of the Universe represents a major theoretical problem in
cosmology, and several approaches have been coined to try and provide
a compelling answer
to this riddle.

The main stream approach is to consider the Universe is filled with
an exotic fluid, known as dark energy
\cite{Padmanabhan:2006ag,Albrecht:2006um,Sahni:2006pa},
but  then one also has to demand the cosmic soup (made of dark energy
and the rest of components) has some Goldilocks properties to comply
with the observations.
Alternatively, the idea
that cosmic acceleration might be due to modifications to general
relativity has received considerable attention as well (see
\cite{Maartens:2007,Durrer:2007re} for reviews and for
specific modifications).
In such frameworks models displaying cosmic acceleration could be
devised with less fine-tuning
and unnaturality as compared to general relativistic dark energy
scenarios \cite{Padmanabhan:2007xy}. Speculations in the direction of
modified gravity are, in principle,
legitimate
as there are no cosmological tests probing scales as large as the
Hubble radius. We only  have reasonable evidence of the validity of
the gravitational inverse square law up to $300$ Mpc (through the ISW
effect) \cite{Peebles:2004qg}. However, the Hubble radius is two
orders of magnitude larger, so our large-scale
tests on general relativity are not stringent enough.

The additional degrees of freedom of these various settings, as
compared to the standard picture of cosmology prior to the revolution
ignited in 1998, have given rise to a collection of new cosmological
evolutions with bold features, future singularites being the most
perplexing  ones. In this respect, attempts to classify somehow the
sort of future singularities to be expected in new devised
cosmological evolutions are of interest.
A popular classification route in the literature \cite{Nojiri:2005sx}
relies exclusively on properties of the curvature tensor
and scalar quantities derived from it. From that perspective, a
number of new terms in cosmology,  such as the celebrated  ``big
rip'' \cite{Caldwell:2003vq}, have been coined to designate
extremality events  associated with blow-ups of scalars constructed
from the curvature tensor, along with less popular ones like
``quiescent singularities'' \cite{Andersson:2000cv}, ``sudden
singularities''   \cite{Barrow:2004xh}, ``big brake'' \cite{brake} or
``big freeze'' \cite{freeze} (the number of names is larger than the
actual name of different
extremality events).

Now, even though treatments of singularities in the fashion of
\cite{Nojiri:2005sx}  are of interest, there are
subtle and most relevant properties inherent to cosmic evolution
which can only be unveiled  through the more sophisticated
consideration of observers (see \cite{suddenferlaz,puiseux} for a
detailed account).  Indeed, curvature is a static concept, as it is
only provides information of what happens at each event. Conversely,
information retrieved
from tracking the observers along their trajectories is more
dynamical in
nature, and therefore more enlightening if carefully analyzed.
Interestingly, this
scheme allows discussing whether the singularities
encountered are weak or strong. Thus, if one's ultimate goal is to
draw rigorous conclusions about the final fate in the Universe, both
approaches are, in our view, complementary.

In this paper we address the problem of future singularities in
modified gravity cosmologies. We examine carefully the interrelation
between the modifications and the singularities
to be expected, and we try and give a unified vision by reporting our
results using the scheme concerned with the behaviour  of curvature
scalars \cite{Nojiri:2005sx} and the one grounded on observers
\cite{suddenferlaz,puiseux}.

Ideally, modifications of general relativity should be derivable of a
parent theory allowing for a covariant formulation of full-fledged
field equations, otherwise, neither density perturbations nor solar
system predictions could be computed. This is, actually,  an aspect
of the problem which does not affect our discussion, as we only work
at the level of the Friedmann and energy conservation equations.
Whenever the literature offers relevant examples for which the
underlying theory is known, we will use
them to illustrate our findings, but, occasionally, we will also
resort to phenomenological examples.

The plan of the paper is as follows. We propose a perturbative
formulation of the Friedmann equations, for which two cases are
distinguished depending on whether there is a critical energy density
(which affects the form of the formulation). Then, we calculate the
corresponding asymptotic expression of the scale factor, and bulding
on
earlier works we present our classification. We round up the
dissertation with relevant examples and summarize in
the last section.

\section{Modifications of Friedmann equation}

There have been many attempts  to modify
Einstein's theory of gravity from different points of view in order
to cope with the observed
acceleration of the expansion of the universe. One possibility arises
from modifications to the Einstein-Hilbert action leading to
the so called $f(R)$ gravity theories (many aspect of this
theoretical setup have been recently reviewed in  \cite{salvc}. The
equations governing the large-scale geometry of the Universe in such
settings
are of fourth order in the metric approach, and, on top of that,  for
$f(R)$ gravity theories to evade compatibility issues with
observational tests complicated models are required \cite{fRevade}
(see however \cite{disc} for a different perspective). 
Mild applications of Ockham's razor principle, combined in graceful
cases with physical motivations, have lead to the consideration that
contending modified gravity schemes could perhaps be more advisable.
This is the case of the proposals originated by assuming the Universe
is a 3-brane
embedded in a higher dimensional bulk.

Instead of grounding our discussion in specific theoretical
frameworks, we propose a perturbative expression
for the Friedmann equation of an expanding universe, which intends to
comprise most of the models
in the literature.

With this aim in mind, we write a modified Friedmann equation in the
form
\begin{equation}
\left(\frac{\dot a}{a}\right)^{2}=H^{2}=h_{0}(\rho-\rho_{*})^{\xi_{0}}
+h_{1}(\rho-\rho_{*})^{\xi_{1}}+\cdots.\label{pert}
\end{equation}
Thus, we assume the squared Hubble factor can be expressed as a power
series in the density $\rho$ of the matter content of the
universe around a specific value $\rho_{*}$, for which a qualitative
change of behavior is expected. The exponents $\xi_{i}$ are real
and ordered, $\xi_{0}<\xi_{1}<\cdots$. The coefficient $h_{0}$ is
obviously positive.

The equation system is closed by assuming, in addition, the validity
of the usual energy conservation equation
\begin{equation}
\dot \rho+ 3H(\rho+p)=0.\label{cons}
\end{equation}

The perturbative formulation represented by Eq. (\ref{pert}) can
accommodate the Friedmann equations of
the existing modified gravity proposals with a known parent covariant
theory, as well as others
with a phenomenological origin. Note, as well, that the
$\Lambda$-cold dark matter (LCDM) or cosmic concordance scenario
\cite{Ostriker:1995rn}
is trivially comprised within this framework:
\[H^{2}=h_{0}+h_{1}(\rho-\rho_{*}),\]
with $\xi_0=0$, $\xi_1=1$, $h_{1}=8\pi G/3$, and
$h_{0}-h_{1}\rho_{*}=\Lambda /3$, so, actually,
the parameter $\rho_{*}$ is not fixed.

The main purpose of the modifications is to provide an accelerated
evolution of the Universe without resorting to an exotic fluid, so it
is usually assumed
the Universe is simply filled  with cold dark  matter
($p=0$), and this will be our working hypothesis as well. In this
case, the energy conservation equation (\ref{cons})
can be straightforwardly integrated:
\begin{equation}
\frac{\dot \rho}{\rho}=-3\frac{\dot a}{a} \Rightarrow \rho a^3=K,
\end{equation}
which gives a one-to-one map between the energy density and the scale
factor through the integration constant
$K$.

If we perform a power expansion of the scale factor in time,
\[    a(t)=c_{0}|t-t_{0}|^{\eta_{0}}+c_{1}|t-t_{1}|^{\eta_{1}}+\cdots,
\]
where the exponents $\eta_{i}$ are real and ordered,
$\eta_{0}<\eta_{1}<\cdots$, following \cite{puiseux} we shall be able
to classify the singularities encountered at a time $t_{0}$. It is
expected that most models allow this sort of expansion. However, there
are models arising in loop quantum cosmology \cite{wands} which show
accelerated oscillations that fall out of this scheme, though most of
our conclusions can be extended to them.


\begin{table}
   \begin{tabular}{|c|c|c|c|c|c|}
   \hline
   ${\eta_{0}}$ & ${\eta_{1}}$ & $\eta_{2}$ &\textbf{Tipler} &
   \textbf{Kr\'olak} & \textbf{N.O.T.} \\
   \hline\hline
   $(-\infty,0)$ & $(\eta_{0},\infty)$ &   $(\eta_{1},\infty)$ &
   Strong & Strong  & I\\ \hline
   $0$ & $(0,1)$ &   $(\eta_{1},\infty)$ &   Weak & Strong & III \\
     \cline{2-6} & $1$ & $(1,2)$ & Weak & Weak & II \\
     \cline{3-6} &  & $[2,\infty)$ & Weak & Weak & IV \\
     \cline{2-6} & $(1,2)$ &  $(\eta_{1},\infty)$ & Weak & Weak & II \\
     \cline{2-6} & $[2,\infty)$ &   $(\eta_{1},\infty)$ & Weak & Weak &  IV \\
   \hline
   $(0,\infty)$ & $(\eta_{0},\infty)$ &   $(\eta_{1},\infty)$ &
Strong & Strong & Crunch\\
   \hline
   \end{tabular}
\caption{ Singularities in FLRW cosmological models}
\end{table}

The classification of singularities in weak and strong follows the
ideas of Ellis and Schmidt \cite{ellis}: The curvature may be finite
or infinite at one event, but what is physically relevant is whether
free-falling (or even accelerated) observers meet the singularity in
finite proper time \cite{HE}. It is clear that if they take infinite time in
reaching the curvature singularity, this would be indetectable.

Furthermore, if instead of ideal unextended observers we consider
finite objects, the key issue is whether tidal forces at the
singularity are \emph{strong} enough to destroy them or \emph{weak},
so that there could be objects that would survive beyond the
singularity. This would mean that the weak singularity is by no means
the end of the universe.

Following this ideas, Tipler \cite{tipler} modelled extended objects
by three
perpendicular vorticity-free Jacobi fields travelling along a causal
geodesic and forming and orthonormal frame with the velocity $u$ of this.
If the geodesic hits a singularity in finite time and the volume
spanned by a set of three three such vectors remains finite, the
singularity is
considered \emph{weak}, since an object is not crushed. Tipler
argues that in this case the metric could be generically extended
beyond the singularity. Otherwise, if the volume is not finite for
every set of vectors, the singularity is considered \emph{strong}.

Thinking of cosmic censorship conjectures, Kr\'olak \cite{krolak}
suggested and alternative definition of strong curvature
singularities that just required diminishing, instead of vanishing,
volume of the finite object and is therefore easier to comply.

Compliance with these definitions for FLRW models can be checked
resorting to integrals of the Ricci tensor along causal geodesics with
respect to proper time $\tau$ \cite{clarke}:

If this integral diverges at a value $\tau_{0}$
\begin{equation}\label{suftipler1}
\int_{0}^{\tau}d\tau'\int_{0}^{\tau'}d\tau''R_{ij}u^{i}u^j,
\end{equation}
the causal geodesic meets a strong curvature singularity according
to Tipler's definition.

And for Kr\'olak's definition divergence of this other integral at $\tau_{0}$
\begin{equation}\label{sufkrolak1}
\int_{0}^{\tau}d\tau'R_{ij}u^{i}u^j,
\end{equation}
means meeting a strong curvature singularity.

The application of these results to FLRW models is summarized in Table
I, which is a simplified version of the one in \cite{suddenferlaz}.

The last columns refers to the classification of future singularities
in \cite{Nojiri:2005sx}):

\begin{itemize}
   \item  Type I: ``Big Rip'': divergent $a$, $\rho$, $p$.

   \item  Type II: ``Sudden'': finite $a$, $\rho$, divergent $p$.

   \item  Type III: ``Big Freeze'': finite $a$, divergent $\rho$, $p$.

   \item  Type IV: ``Big Brake'': finite $a$, $\rho$, $p$, but
   divergent higher derivatives.
\end{itemize}

Even though the modifications will only lead to acceleration for
certain values of the parameter $\xi_0$,
our forthcoming discussion on the asymptotic behavior of the scale
factor is valid for any value of $\xi_0$, and it only relies
on the ordering of the exponents.  For this reason, our scheme
comprises as well modifications to gravity which are not able to
explaining the current
acceleration, such as, for instance, the non-self-accelerating branch
of the DGP modification scenario.

Inserting the modified Friedmann equation into the conservation
equation one gets:
\begin{eqnarray}
\frac{\dot\rho}{\rho}&=&
-3\sqrt{h_{0}}(\rho-\rho_{*})^{\xi_{0}/2}-
\nonumber\\&-&
\frac{3}{2}\frac{h_{1}}{\sqrt{h_{0}}}(\rho-\rho_{*})^{\xi_{1}-\xi_{0}/2}
+\cdots\quad\,\label{conspert}.
\end{eqnarray}

Our purpose it to integrate the latter by considering all the
possibilities
which arise from different values of the parameters, and then use the
aforementioned map
between the energy density and the scale factor so that we can
finally obtain asymptotic
expressions for the expansionary behaviour of the models. Then, we
will identify the
specific late-time behaviour of the models, focusing on the existence
of future singularities of various types. This classification resorts
to earlier
works by ourselves.

A  separate treatment of the cases $\rho_{*}=0$ and $\rho_{*}\not=0$
cases is required, so we split the discussion into
two subsections.
\subsection{Absent critical density}

In the case of a theory with no critical density, i.e. 
density
$\rho_{*}=0$, expressions get considerably simplified:
\begin{eqnarray}\frac{\dot\rho}{\rho}&\simeq&-3\sqrt{h_{0}}\rho^{\xi_{0}/2},\nonumber\\
\rho(t)&\simeq& \left\{\begin{array}{ll}
\left\{\displaystyle\frac{3}{2}\xi_{0}\sqrt{h_{0}}(t-t_{0})\right\}^{-2/\xi_{0}}
&\textrm{for }\xi_{0}\neq 0,\\\\\displaystyle
e^{-3\sqrt{h_{0}}(t-t_{0})} &\textrm{for } \xi_{0}=
0.\end{array}\right.
\end{eqnarray}
Correspondingly, in terms of the expansion factor we get
\[a(t)\simeq \sqrt[3]{K}
\left\{\frac{3}{2}\xi_{0}\sqrt{h_{0}}(t-t_{0})\right\}^{2/3\xi_{0}},\]
which provides the following expected results:
\begin{itemize}
   \item  $\xi_{0}<0$: As matter density decreases smoothly, an
eventual blow up of the corrections to the Friedmann equation is
approached.  At a
finite time $t_{0}$ the scale factor
   becomes infinite, and the Universe experiences a  type of
singularity  which has been called ``big rip''
   \cite{Caldwell:2003vq}
   (type I in the
   classification in \cite{Nojiri:2005sx}).

   \item  $\xi_{0}>0$: The matter density decreases and the scale
   factor increases smoothly as $t$ grows towards infinity. This
case comprises
   both quintessence-like behaviors for $\xi_{0}\in (0,2/3)$, and
non-accelerated
   evolutions for $\xi_{0}\ge 2/3$.

   \item $\xi_{0}=0$: The lowest order term is that of a
   cosmological constant, and we have to resort to
   the first correction with a positive exponent $\xi_{1}$, which
leads  again to
   an expression solvable as a Bernouilli equation:
\begin{eqnarray}\frac{\dot\rho}{\rho}&\simeq&-3\sqrt{h_{0}}
-\frac{3}{2}\frac{h_{1}}{\sqrt{h_{0}}}\rho^{\xi_{1}}\\\nonumber
\rho(t)&\simeq&
\left\{e^{3\xi_{1}\sqrt{h_{0}}(t-t_{0})}-\frac{1}{2}\frac{h_{1}}{h_{0}}\right\}^{-1/\xi_{1}}.
\end{eqnarray}
In this case
\begin{eqnarray}
a(t)&\simeq &\sqrt[3]{K}
\left\{e^{3\xi_{1}\sqrt{h_{0}}(t-t_{0})}-\frac{1}{2}\frac{h_{1}}{h_{0}}\right\}^{1/3\xi_{1}},
\end{eqnarray}
so, this situation represents an exponential expansion of the
Universe, with a
corresponding exponential decrease of matter density, with no
future singularity at all.
\end{itemize}
Therefore,  in the case $\rho_{*}=0$, the modifications considered
do not
produce a qualitative change of behavior towards the future, except
for dramatic modifications produced by negative exponents, which lead
to a
``big rip" singularity in the future.

We close this subsection with several  examples which  fit in
$\rho_{*}=0$ case of the general
perturbative expression of $H^{2}$ we started from, namely they satisfy
\begin{equation}
H^{2}=h_0\rho^{{\xi}_0}+h_1\rho^{{\xi}_1}+\dots \label{examples}.
\end{equation}

These results are summarized in Table II.

\begin{table}
   \begin{tabular}{|c|c|c|c|}
   \hline
   ${\xi_{0}}$ &  \textbf{Tipler} &
   \textbf{Kr\'olak} & \textbf{N.O.T.} \\
   \hline\hline
   $(-\infty,0)$ &
   Strong & Strong  & I\\ \hline
   $[0,\infty)$ &
Non-singular & Non-singular & Non-singular\\
   \hline
   \end{tabular}
\caption{ Singularities in models without critical density}
\end{table}

The first case we consider for illustration is that of the power-law
Cardassian  models \cite{card}, for which
\begin{equation}
H^{2}=\frac{8}{3} G \pi  \rho  \left(1+\left(\frac{\rho }{\rho_{\rm
card}}\right)^{n-1}\right).
\end{equation}
This expression can be accommodated into (\ref{examples})
with the following identifications between our parameters and those
of the original reference: $\xi_0=1$, $\xi_1=n<2/3$,  $h_0={8\pi
G}/3$, $h_1=({8\pi G}/{3})\rho_{card}^{1-n}>0$. The constant
$\rho_{card}$ signals the amount of matter energy density $\rho$
below which the Cardassian corrections start to dominate
($\rho_{card}\sim\rho$)

DGP cosmologies
\cite{Dvali:2000rv,Deffayet:2000uy,Sahni:2002dx,Lue:2004za,Sahni:2005mc}
provide another interesting set of examples. If the brane has no
tension and the bulk is the Minkowski spacetime, one has
\cite{Sahni:2002dx,dgpparam}
\begin{equation}
\frac{H}{H_0}=\sqrt{\Omega_{r_c}+\frac{8 G \pi\rho}{3
H_0^{2}}}\pm\sqrt{\Omega_{r_c}}.
\end{equation}
Here $H_0$ is the value of the Hubble factor today, and $\Omega_{r_c}$
is the present value of the fractional energy density associated
with the scale at which the crossover to a corrections dominated
regime occurs.
In the perturbative formulation required for the discussion we get
\begin{equation}
H^{2}={\Omega_{r_c}} (1\pm1)^{2}H_0^{2}+\frac{8 G \pi  \rho
}{3}(1\pm1)\mp\frac{16 G^{2} \pi   ^{2} \rho ^{2}}{9 {H_0}^{2}
{\Omega_{r_c}}}+\dots .
\end{equation}
The self-accelerating branch \cite{Deffayet:2000uy,Sahni:2002dx}
arises by taking the upper signs, and it is characterized by
$\xi_0=0$ and $\xi_1=1$, whereas
for the so called normal branch \cite{Sahni:2002dx,Lue:2004za},
which arises by taking the lowers signs, one has $\xi_0=2$.

Finally, we can bring about the reinterpretation of  Chaplygin-like
cosmic evolutions as a modified gravity proposal
\cite{Barreiro:2004bd,Chimento:2005au}. In these frameworks one
has
\begin{equation}
H^{2}=\frac{8 \pi G}{3}\left(A+\rho^{\gamma(\alpha
+1)}\right)^{\frac{1}{\alpha +1}}.
\end{equation}
with $\alpha>0$ and $A=(3H_0^{2}/(8 \pi
G))^{(1+\alpha)}\left(1-\Omega_m^{1+\alpha}\right)$, and $\gamma=1$
in the models considered in \cite{Barreiro:2004bd}, whereas in the
more general framework of \cite{Chimento:2005au} $\gamma$ is free.
Two cases are to be distinguished. In the $\gamma>0$ case,
the correspondence up to order $\gamma(1+\alpha)$ in $\rho$ is given
by
$h_0=({8 G\pi}/{3}) A^{{1}/{(1+\alpha)}}$, $\xi_0=0$, $h_1=({8  \pi
G}/(3 (1+\alpha))A^{-\alpha /(1+\alpha)}$,
$\xi_1=\gamma(\alpha+1)$.  But in the $\gamma<0$ the correspondence
is rather different, as the identification up to order $\gamma$
is given by $h_0={8 G\pi}/{3}$ and $\xi_0=\gamma$.

In all the examples but the last one, according to the discussion
above, no singular fate of the universe is faced.
On the  contrary, in the last kind of models the singularity is of
``big rip" type.

\subsection{Non-trivial critical density}
New features appear for general modifications endowed with  a
non-trivial critical
density $\rho_{*}$. For this case, we assume the matter density has
an expansion around the critical value $\rho_{*}=\rho(t_{0})$ at a
time $t_{0}$:
\begin{eqnarray}
\rho(t)&=&\rho_{*}+\rho_{1}(t_{0}-t)^{\tilde\eta_{1}}+\rho_{2}(t_{0}-t)^{\tilde\eta_{2}}+
\cdots\,,
\end{eqnarray}
and from the latter we obtain
\begin{eqnarray}
a(t)&=&\frac{\sqrt[3]{K}}{\rho_{*}^{1/3}}\left(1-\frac{\rho_{1}}{3\rho_{*}}(t_{0}-t)^{\tilde\eta_{1}}+\cdots\right)
,\end{eqnarray}
so that the first exponents are the same in both expansions,
\[\eta_{0}=1,\qquad \eta_{1}=\tilde\eta_{1},\quad\ldots\]
and we may drop the tildes.

At lowest order we have,
\begin{eqnarray*}
\frac{\dot\rho}{\rho}&\simeq&
-3\sqrt{h_{0}}(\rho-\rho_{*})^{\xi_{0}/2}
=-3\sqrt{h_{0}}\left\{\rho_{1}(t_{0}-t)^{\eta_{1}}\right\}^{\xi_{0}/2},
\end{eqnarray*}
which upon the requirement of compatibility with Eq. (\ref{conspert})
fixes the first exponent as
\begin{equation}
\eta_{1}=\frac{2}{2-\xi_{0}}.
\end{equation}
The following three cases are to be distinguished:
\begin{itemize}
   \item  $\xi_{0}<0$: Since
   $0<\eta_{1}<1$, according to \cite{puiseux} or Table I, these
models have a
singularity at $t_{0}$ with divergent $H$ (a ``big freeze'' or
singularity type III
\cite{Nojiri:2005sx}), which is a weak curvature singularity
according to
Tipler \cite{tipler}, but strong according to Kr\'olak \cite{krolak}.

\item $\xi_{0}\in (0,2)$: In this case
$\eta_{1}>1$, so these models could show a weak singularity at
$t_{0}$ according to \cite{puiseux} or Table I (sudden singularity
\cite{Barrow:2004xh} or type II in
\cite{Nojiri:2005sx}, or even type IV if $\eta_{1}\ge 2$, $\xi_{0}\ge
1$).

\item  $\xi_{0}=0$: The cosmological constant term is dominant
against modifications of the Friedmann equation. At first order, we
have
\[\frac{\dot\rho}{\rho}\simeq-
\frac{\eta_{1}\rho_{1}}{\rho_{*}}(t_{0}-t)^{\eta_{1}-1}=-3\sqrt{h_{0}},
\] that is, we find a linear behavior for matter density:
\[\eta_{1}=1,\quad\rho_{1}=3\rho_{*}\sqrt{h_{0}}.\]
This being so, it turns out we have to expand the equation a bit
further in order to reveal
new qualitative behavior:
\begin{eqnarray*}
\hspace{0.8cm}\frac{\dot\rho}{\rho}&\simeq&
-3\sqrt{h_{0}}
-\eta_{2}\frac{\rho_{2}}{\rho_{*}}\sqrt{h_{0}}(t_{0}-t)^{\eta_{2}-1}+\dots\nonumber\\
\hspace{0.6cm}&=&-3\sqrt{h_{0}}-\frac{3}{2}\frac{h_{1}}{\sqrt{h_{0}}}\left\{3\sqrt{h_{0}}\rho_{*}(t_{0}-t)\right\}^{\xi_{1}}+\cdots\\\hspace{0.2cm}&=&-3\sqrt{h_{0}}-\frac{3}{2}\frac{h_{1}}{\sqrt{h_{0}}}(\rho-\rho_{*})^{\xi_{1}}+
\cdots.
\end{eqnarray*}

Necessarily,
\begin{equation}
\hspace{0.9cm}\eta_{2}=1+\xi_{1},\quad
\rho_{2}=\frac{1}{2\eta_{2}}\frac{h_{1}}{h_{0}}\left(3\sqrt{h_{0}}\rho_{*}
\right)^{\eta_{2}},
\end{equation}
and therefore, according to \cite{puiseux} or Table I,
\begin{eqnarray}
\hspace{0.6cm}a(t)&=&\frac{\sqrt[3]{K}}{\rho_{*}^{1/3}}\left\{1-\frac{\rho_{1}}{3\rho_{*}}(t_{0}-t)-
\frac{\rho_{2}}{3\rho_{*}}(t_{0}-t)^{\eta_{2}}
+\right.\nonumber\\
\hspace{0.4cm}&&
\left.
\cdots\right\},\end{eqnarray}
there is a singularity at $t_{0}$ due to the lack of smoothness of
the density and the scale factor. But this singularity is weak
in  both Tipler's \cite{tipler} and Kr\'olak' \cite{krolak}
classification, so  it
does not exert any infinite distortion on finite objects going
through it
and cannot, therefore, be considered as a  final stage of the
Universe. It
is a sudden singularity or type II in \cite{Nojiri:2005sx}) for which
the
scale factor and the density remain finite, but $\dot H$ blows up.

It is worthwhile mentioning that milder singularities for which $H$
and also $\dot H$ are finite (type IV in \cite{Nojiri:2005sx}) could,
in principle,
appear within this framework, but they would involve  choosing
$\eta_{2}\ge2$, and thereby $\xi_{1}\ge1$, so that the linear
term in the density in Friedmann equation would be absent.
\end{itemize}

Obviously models with analytical expansion, that is, natural exponents
$\xi_{0}$, $\xi_{1}$,\ldots (such as LCDM, for instance) do not show
future singularities, neither weak nor strong.

These results are summarized in Table III.

\begin{table}
   \begin{tabular}{|c|c|c|c|c|}
   \hline
   ${\xi_{0}}$ & ${\xi_{1}}$ & \textbf{Tipler} &
   \textbf{Kr\'olak} & \textbf{N.O.T.} \\
   \hline\hline
   $(-\infty,0)$ & $(\xi_{0},\infty)$ &
   Weak & Strong  & III\\ \hline
   $0$ & $(0,1)$ &      Weak & Weak & II \\
   \cline{2-5} & $[1,\infty)$ &  Weak & Weak & IV \\
   \hline
   $(0,1)$ & $(\xi_{0},\infty)$ &
Weak & Weak & II\\ \hline
   $[1,2)$ & $(\xi_{0},\infty)$ &
Weak & Weak & IV\\
   \hline
   \end{tabular}
\caption{ Singularities in models with critical density}
\end{table}

The normal branch of DGP cosmologies provide a relevant example for
this
section. If the bulk on which the brane lives is an anti-de Sitter
spacetime one has
\begin{equation}
\frac{H}{H_0}=\sqrt{\frac{8 G \pi  \rho -|\Lambda_b |}{3
H_0^{2}}+\Omega_{r_c}}-\sqrt{\Omega_{r_c}}.
\end{equation}
The identification with our perturbed formulation is given by
$\rho_{*}=(\vert\Lambda_{b}\vert-3 H_0^{2} \Omega_ {r_c})/({8  \pi G
})$,
$h_0=\Omega_{r_c}H_{0}^{2}$, $\xi_0=0$, $h_1=-4H_0\sqrt{2 \pi
G\Omega_{r_c}}/(\sqrt{3})$ and $\xi_1=1/2$.
This singularity is a sudden one, also referred to as quiescent
\cite{singbrane}, or  using our terminology, it is a weak
extremality event. A slight variation leading to a singularity of the
same sort consists in letting
the brane have a negative brane tension $\sigma$. In this case the
bulk can either be the Minkowski or
the anti-de Sitter spacetime. The above expression can be adapted to
this variation by simply letting
$\rho\to\rho+\sigma$ and $\rho_{*}\to\rho_{*}+\sigma$.

Finally, we may consider models arising in loop quantum cosmology as
those in \cite{battisti}, for which \[H^{2}=\frac{8\pi
G}{3}\rho\left(1- \frac{\rho}{\rho_{*}}\right),\] but for these
models the critical density is relevant for the high density regime,
imposing a maximum density which is reached as the energy density grows.
As this is the opposite of our working hypothesis (remember we demanded $\dot\rho<0$) these models do not
quite fit in our description here, but could be treated in an analogous way,
with the corresponding adjustements.

\section{Discussion \label{discuss}}
We here put forward a detailed classification of the future behavior
of
FRW cosmologies in modified gravity proposals. Departures from the
standard description of the expansion of the Universe according to
Einstein's theory
have been considered of interest, as they could provide an
explanation of what is
the agent responsible for the accelerated expansion of the Universe.

The main question we pose is what are the characteristics of the
modifications in connection with the
presence of a singular future behavior of the Universe. As we have
reflected here, not all the
relevant properties of cosmic evolution emerge by considering
curvature scalars, and the deeper insight
provided by the consideration of observers is needed.

The spirit of the modified gravity proposals we consider is to assume
the Universe is simply filled with cosmic dust,
and no blueshifting component whatsoever is considered (unlike when
one assumes the current cosmic acceleration
is due to an exotic fluid or dark energy). Our starting point is a
perturbative low-energy or infrared
expansion of  the modified Friedmann equation. Two classes emerge:
those with a critical energy density
and those without it. We find one has to consider at most the
exponents of the first two terms of the expansion
in order to differentiate the possible behaviors, and, more
importantly, whether the future singularity, if it exists,
is weak or strong.

The scheme we propose provides an easy route to conclude the sort of
singular behavior present
in potential new  candidates to  explain the current acceleration in
the universe in terms of  a modification
of gravity. The classification we put forward is complementary to
others, but provide a deeper insight and allow
an important further degree of refinement.

\section*{Acknowledgments}L.F.-J. is supported by the Spanish Ministry
of Education and Science Project FIS-2005-05198.  R.L. is supported by
the University of the Basque Country through research grant
GIU06/37 and by the Spanish Ministry of Education and
Culture through research grant FIS2007-61800. L. F.-J. wishes to
thank the University of the Basque
Country for their hospitality and facilities to carry out this work.

\end{document}